\documentstyle[aps,preprint,tighten]{revtex}
\def\p{\phi}
\def\P{\Phi}

\def\ps{\psi}
\def\a{\alpha}

\def\g{\gamma}
\def\d{\delta}
\def\D{\Delta}
\def\i{\iota}
\def\be{\begin{equation}}
\def\ee{\end{equation}}
\def\bea{\begin{eqnarray}}
\def\eea{\end{eqnarray}}
\def\nn{\nonumber}
\def\x{\tilde{x}}
\def\s{\sigma}

\def\l{\lambda}
\def\L{\Lambda}
\def\m{\mu}

\def\ph{\phantom}

\def\H{{\cal{H}}}

\def\sinh{\textrm{sinh}}
\def\mod{\textrm{mod}\ }

\def\dim{\textrm{dim}\ }
\def\det{\textrm{det}}
\def\log{\textrm{ln}}
\def\tr{\textrm{tr}}
\def\no{\noindent}
\def\CC{{\rm\kern.24em \vrule width.04em height1.46ex depth-.07ex
\kern-.30em C}}
\def\RR{{\rm
         \vrule width.04em height1.58ex depth-.0ex
         \kern-.04em R}}
\def\id{{\rm 1\kern-.22em l}}
\def\1{\textrm{\bf{1}}}
\def\ra{\rightarrow}
\begin{document}
\title{On the Spectrum of the XXZ-chain at roots
of unity}
\author{ Daniel Braak\footnote{e-mail: braakd@physik.uni-augsburg.de}}
\address{Institut f\"ur theoretische Physik II, Universit\"at Augsburg,\\
86135 Augsburg, Germany}
\author{Natan Andrei\footnote{email: natan@physics.rutgers.edu}}
\address{Material Theory Center, Department of Physics and Astronomy, Rutgers University,\\
Piscataway, NJ 08855, USA}
\date{June 11, 2001}
\maketitle
\begin{abstract}
In a recent paper (cond-mat/0009279),  Fabricius and McCoy
studied the spectrum of the spin 1/2 XXZ model at roots of unity,
i.e. $\D=(q+q^{-1})/2$ with $q^{2N}=1$ for integer $N\ge 2$. They
found
 a certain pattern of degeneracies  and linked it
to the $sl_2$-loop symmetry present in the  commensurable
spin sector - $S^z\equiv 0\ \mod N$.

We show that the degeneracies are due to
an unusual type of zero-energy ``transparent'' excitation, the
{\it cyclic bound state}. The cyclic bound states exist both
in the commensurable and in the incommensurable sectors indicating a
 symmetry group present, of  which  $sl_2$-loop algebra is
a partial manifestation. Our approach treats both sectors on even footing and
allows us to obtain analytically an explicit expression for the degeneracies
in the case $N=3$.
\end{abstract}
\noindent

\section{Introduction}
In a recent interesting work \cite{mcc1,mcc2} the spin - 1/2 XXZ model,
\be
H=\sum_{j=1}^L \s_j^+\s_{j+1}^-+\s_j^-\s_{j+1}^+ + (\D/2)\s_j^z\s^z_{j+1},
\label{ham}
\ee
was studied on a chain with $L$ sites and periodic boundary conditions ($\s_{L+1}=\s_1$)
for the special values of the anisotropy.
\be
\D = \frac{1}{2}(q+q^{-1}) \ \ \ {\textrm{with}}\ \ q^{2N}=1,
\label{rootofunity}
\ee
 A large  symmetry algebra was shown to be present, manifesting
itself in a rich pattern of degeneracies.

For $N=2$, $\D=0$ the study was carried out analytically.
 The model reduces to the XY model, equivalent to free
fermions. The degeneracies, in this case, are  due
to ``pairs'' of spin-down excitations with momenta $p_{1,2}$,
 satisfying the condition
\[
p_1+p_2=\pi \; \mod 2\pi,
\]
because the energy of the ``pair''  is zero:
\[
\cos p_1 +\cos p_2 =0.
\]
A combinatorial argument \cite{mcc1} then  shows that an eigenstate
with $S^z=S^z_{max}$
is degenerate with states having different spin $S^z=S^z_{max}-2l$,
$0\le l\le S^z_{max}$.
The dimension of the multiplet with $S^z=S^z_{max}-2l$ is
(assuming $L$ to be even):
\be
{S^z_{max}\choose l},
\label{deg2even}
\ee
if $S^z_{max}$ is even, otherwise it is
\be
{S^z_{max}\pm 1\choose l},
\label{deg2odd}
\ee
with the $(\pm)$ - sign  depending on the  $ S^z_{max}$ ``parent state''.
Whereas the formula,  eqn (\ref{deg2even}), for the commensurable case (2 divides $S^z_{max}$), is independent of the parent state, the formula
for the incommensurable
case, eqn (\ref{deg2odd}), although similar, does depend on it.

Deguchi {\it et al.} \cite{mcc1} relate this degeneracy to the
$sl_2$-loop algebra, which is a symmetry of the Hamiltonian for $\D=0$
and $S^z_{max}$ even. This symmetry is not realized in the
case $S^z_{max}$ odd, but in \cite{mcc1} it is argued that a
certain residual of it is still responsible for the degeneracies
given by (\ref{deg2odd}).

The degeneracies for $N\ge 3$ were studied numerically
in \cite{mcc1}
and the authors found a surprisingly simple generalization
of (\ref{deg2even}),(\ref{deg2odd}): The state with $S^z=S^z_{max}$
is degenerate with states having $S^z=S^z_{max}-lN$, and the
corresponding multiplet has dimension,
\be
{2S^z_{max}/N \choose l}
\label{degNdiv}
\ee
in the commensurable case,
\be
S^z_{max} \equiv 0\ \mod N,
\ee
and
\be
{2[S^z_{max}/N]+\alpha \choose l}
\label{degNndiv}
\ee
in the incommensurable case, $N$ does not divide $S^z_{max}$ ($[x]$ is
the Gauss step function and $\alpha$ may take values 0,1 or 2).

 \bigskip

In this paper we show that the  degeneracies arise due to the presence
of  multi-particle  {\it transparent} excitations which
are present in the spectrum
 for $q$ at a root of unity. Transparent excitations
 are spin carrying, zero-energy
excitations that can be added
to an eigenstate, without changing its Bethe-Ansatz parameters, leading
this way to degenerate multiplets, manifestation of some underlying symmetry.
As this  symmetry patterns are   present in the full Hilbert space
the symmetry cannot be identified with
the  $sl(2)$-loop,  valid only in commensurable sectors.
Our approach allows us to derive an analytic expression
for the multiplet degeneracies and, in particular in the incommensurable
case, identify
what feature of the parent state leads to
the various degeneracies.

Our paper is organized as follows. In section II we discuss
the $SU(2)$ and  quantum group $U_q(sl_2)$ symmetries in the
 spin-1/2 chains and give
an interpretation of these symmetries within the Bethe ansatz language
in terms of single-particle {\it transparent} excitations.

In section III we introduce a class of  multi-particle {\it transparent}
excitations.
 They are allowed only when
the anisotropy of  the XXZ-chain has property
(\ref{rootofunity}), and lead to the multiplets described
by (\ref{degNdiv}) and (\ref{degNndiv}). We give the general construction
and properties of these states for all $N$. Then we specialize to the
simplest nontrivial case $N=3$ and derive formula (\ref{degNdiv})
for $l=1$ while giving an  argument for its validity 
for $l>1$. 
In the incommensurable case we show that three cases arise when the
degeneracy of a multiplet built on a $S^z_{max}$
parent state is considered, these cases corresponding to the presence in it
 of the special single-particle  excitations, considered in Section II.

The appendix gives details of the derivation and some numerical examples,
which indicate that most but not all degeneracies of the spectrum
can be explained within this framework.

\section{transparent excitations and symmetries of the spin chain}
A spin state
 in the sector with fixed $S^z\ge 0$
can  be written as,
\be
|\Psi\rangle =\sum_{1\le n_1<\ldots n_M\le L}f(n_1,\dots,n_M)\s^-_{n_1}
\ldots\s^-_{n_M}|0\rangle,
\label{bastate}
\ee
with $M=L/2-S^z$ and the reference state $|0\rangle$ having all spins up.
 For eigenstates of (\ref{ham}) the coefficients $f(n_1,\ldots,n_M)$ take
a Bethe ansatz form:
\be
f(n_1,\ldots,n_M)=\sum_{P\in S_M}A_P\exp\left(i\sum_j^Mk_{P(j)}n_j\right),
\label{wavef}
\ee
 parameterized by  $M$ numbers, the spin momenta $k_j$, $j=1\ldots M$.
Imposing periodic boundary conditions  the
$k_j$ are determined by:
\be
e^{-ik_jL}=\prod_{l\neq j}^MS(k_j,k_l),
\label{ba}
\ee
where the $S$-matrices $S(k_j,k_l)$ are:
\be
S(k_j,k_l)=-\frac{e^{i(k_j+k_l)}+1-2\D e^{ik_l}}{e^{i(k_j+k_l)}
+1-2\D e^{ik_j}}.
\label{sma}
\ee
The $S(k_j,k_l)$  are determined from (\ref{ham}) and relate
the amplitude
in the wave function for the  down-spin (magnon)
associated with $k_j$ to be to the left
of the magnon  associated with $k_l$,
 to the amplitude with their order reversed.

Introducing the $\l$ - parameterization via,
\be
e^{ik_j}=-\frac{\sinh\frac{\g}{2}(\l_j+i)}{\sinh\frac{\g}{2}(\l_j-i)}
\label{l-param}
\ee
with $\g$ defined  by $\D=(q+q^{-1})/2=-\cos\g$, $-1 < \D < 1$, the S-matrices take the form,
\be
S(\l_j-\l_k)=\frac{\sinh \frac{\g}{2}(\l_j-\l_k-2i)}
{\sinh \frac{\g}{2}(\l_j-\l_k+2i)}.
\label{smal}
\ee
For the case $\D=1$, the XXX model, we have instead
\be
e^{ik_j}=\frac{\l_j+i}{\l_j-i}
\ee
and
\be
S(\l_j-\l_k)=\frac{\l_j-\l_k-2i}{\l_j-\l_k+2i}.
\label{smaxxx}
\ee

Having characterized the states we proceed to study the manifestations
of symmetry in the spectrum, relating it to the presence of
{\it transparent} excitations -
a concept which we shall significantly generalize as we progress. Begin
with the XXX - chain ($\D=1$) which is $SU(2)$ invariant.
The BA equations (\ref{ba})  determine the
parameters $\l_1,\ldots \l_M$ under the assumption that all of
them are finite. What happens, if some of the $\l$ diverge?
For $\l_0\ra \infty$, the $S$-matrix (\ref{smaxxx}), $S(\l_0,\l_k)\ra 1$,
independently of $\l_k$, and the corresponding factors drop out
of (\ref{ba}). That means, given a set of finite $\l_1\ldots \l_M$,
which describe an eigenstate with $S^z_{max}=L/2-M$,
one can generate an eigenstate with $S^z= S_{max}^z-l$ by adding $l$
spin down excitations with parameter $\l=\infty$.
The physical interpretation of these ``singular'' values of the
$\l$'s is very simple: They are nothing but elementary excitations
(``magnons'') with momentum $k_0=0$ and energy $E=0$. These states
have $S$-matrix $S(0,k_j)=1$ with all other particles and among themselves,
they are therefore {\it transparent},
which is the reason that more than one of them can appear in an eigenstate
of (\ref{ham}). (For nonzero $k_j$, the $S$-matrix $S(k_j,k_j)=-1$,
forbidding more than one excitation having momentum $k_j$).
By ``adding'' we mean the following operation:
Given the amplitudes $A_P$ in (\ref{wavef}) we construct the
components of the wavefunction
$f(n_1,\ldots,n_M,n_{M+1},\ldots,n_{M+l})$ through,
\be
f(n_1,\ldots,n_M,n_{M+1},\ldots,n_{M+l})=\sum_{Q\in S_{M+l}}A_{P(Q)}
\exp\left(i\sum_j^{M+l}k_{Q(j)}n_j\right).
\label{wavef2}
\ee
Here $P(Q)$ denotes a permutation $\in S_M$, such that for $1\le j \le M$:
\be
P^{-1}(j)=\textrm{number of}\ m\in \{1\ldots M\}\ \textrm{for which}\
Q^{-1}(m)\le Q^{-1}(j).
\label{permp}
\ee
and  $k_l=0$ for $l>M$. (An example of the construction is given
in Appendix A.)
This way we may construct the explicit Bethe wavefunction of
a state containing $l$ of the zero-momentum excitations, given
an eigenstate (\ref{bastate}) of the hamiltonian described
by momenta $\{k_j\}$ and
amplitudes $\{A_P\}$.

 The new states can be generated directly
from the state (\ref{bastate}) by acting on it $l$ times with the operator
\be
\hat{S}^-=\sum_{j=1}^L\s_j^-,
\label{genxxx}
\ee
which coincides with the spin lowering operator
in the $L$-fold product of spin 1/2 representations.
The global $SU(2)$ symmetry corresponds therefore to the existence
of transparent excitations with zero energy. At the same time, this
leads to singular parameters in the Bethe ansatz.

Now examine the case $\D\neq 1$, i.e. $q \neq 1 $. Here the situation
 is more interesting.
Introducing $k_q$ via  $e^{ik_q}=q$, we note that the  momentum  $k_q$
particle  has zero energy and interacts
(see eqn (\ref{sma})) with other
particles  $k_j$ via an S-matrix,
\be
S(\pm k_q,k_j)= q^{\mp 2},
\label{smatrixq}
\ee
that is independent of $k_j$. However $S(\pm k_q,k_j) \ne 1$, so
 this excitation is  {\it not} transparent.
The independence of the $S$-matrix
on the momentum $k_j$ of the second particle makes the
excitation with $\pm k_q$ nevertheless
 a  candidate for a quantum symmetry,
which would operate on a state without affecting the
parameters $\{k_j\}$ of that state as in (\ref{wavef2}).
Moreover, as numerator and denominator of $S(k_q,k_q)$ vanish,
the $S$-matrix of these states among themselves can be chosen to be
unity and therefore more than one of them can appear in a given state.
This is not the case for  periodic BC where the parameters  $\{k_j\}$ of
 the parent state  are
affected by adding $l$ excitations with $\pm k_q$. The Bethe-Ansatz equation
now are:
\bea
e^{-ik_jL}&=&\prod^M_{l\neq j}S(k_j,k_l)q^{\pm 2l}
\label{qba1}\\
q^{\pm L} &=& q^{\pm 2M},
\label{qba2}
\eea
and  not only lead to a shift of the parameters $k_j$ of the
parent state
but they are consistent only at special values of $S^z=L/2-M$ and $q$:
$q^{L-2M}=1$. Therefore the XXZ model with periodic BC does not possess this
symmetry in general.\footnote{The $k_j$ are shifted by a constant, this is
 equivalent to a boundary twist. Adding a boundary term \cite{pas} to
the hamiltonian with open BC,
the excitation with $k_q$ becomes allowed:
The boundary field is constructed in a way that
 this state (but not the state with $-k_q$!) is not
reflected at the boundary and therefore we do not have (\ref{qba2}).
Moreover, in the open boundary equation analogous  to (\ref{qba1}),
 the $q$-dependent factor
drops out and the parameters of the parent state are unchanged. The
$U_q(sl_2)$ symmetry is then manifest.}

However under certain conditions, the states with $\pm k_q$, can exist on a periodic
chain. Assume that $q^{2N}=1$. Then   adding $mN, \ m=1,2 \dots $
excitations with momentum
$k_q$ leaves the original equations unmodified (equation
(\ref{qba1}) coincides with (\ref{ba})), because the $N$ particles
{\it together} are again transparent.
But we have still equation (\ref{qba2}), which entails that
\be
S^z\equiv 0\  \mod N.
\label{commc}
\ee
That means, that in certain sectors of the Hilbert space, satisfying
the commensurability condition (\ref{commc}), the excitation
containing $N$ of the elementary ones has zero energy and is
transparent to all other excitations. Together with other transparent
excitations, to be discussed in the next section, it induces
 a quantum symmetry (that means, it has no classical counterpart
as for the XXX - chain).

In terms of the $\l_j$-parameterization, the $k_q$-state is again singular:
$\l_q=\infty$, rendering this parameterization useless in dealing
with the $q$-symmetry. The formula for $f(n_1,\ldots,n_{M+1})$ is now
more complicated than (\ref{wavef2}), because $k_q\neq 0$ and the
$S$-matrices (\ref{smatrixq}) are nontrivial. Nevertheless the state
can be constructed, because its BA equations are the
same as for the parent state. Further, there is  an analogue to
(\ref{genxxx}), namely,
\be
\hat{S_q}^-=\sum_{j=1}^Lq^{\s^z/2}\otimes\cdots \otimes\s_j^-\otimes\cdots
\otimes q^{-\s^z/2},
\label{genq1}
\ee
which is the quantum deformed version of the spin lowering
operator, belonging to $U_q(sl_2)$.
The noncommutative coproduct structure of $U_q(sl_2)$ is just well suited to
generate the wavefunction
having an excitation with $k_q$
in accord with $S(k_q,k_j)=q^{-2}$. The isomorphic
representation \cite{jim},
\be
\hat{T_q}^-=\sum_{j=1}^Lq^{-\s^z/2}\otimes\cdots \otimes\s_j^-\otimes\cdots
\otimes q^{\s^z/2},
\label{genq2}
\ee
likewise creates a state with $-k_q$.

While  these are not symmetry operations, we saw that exciting $N$ $k_q$'s
or  $-k_q$'s does induce symmetry. The generators for both of these are,
\be
S^{-(N)}=\frac{(\hat{S_q}^-)^N}{[N]_q!},\ \ \
T^{-(N)}=\frac{(\hat{T_q}^-)^N}{[N]_q!}
\label{sl2loopgen}
\ee
as follows  from eqns (\ref{genq1},\ref{genq2}).

In \cite{mcc1} it was shown that $S^{\pm(N)},T^{\pm(N)}$
 together with $S^z$ generate  the
$sl_2$-loop algebra, which is therefore a symmetry in the commensurable
sectors of the periodic XXZ model at $q^{2N}=1$.
However, the incommensurable sectors do {\it not}
have this symmetry
because the excitations with $\pm k_q$ violate the periodic BC
\cite{mcc1}.

In the following section we shall show that the excitations
generated by $S^{-(N)},T^{-(N)}$ form  part of a  larger set
of transparent multi-particle excitations, the {\it cyclic bound states},
existing in the XXZ model at roots of unity. Taking all of them into
account, we can derive (\ref{degNdiv}) and find  analogous formulae
in the incommensurable case where $N$ does not divide $S^z_{max}$.

\section{Bound states and cyclic bound states}

In this section we show there exists a large class of transparent
excitations both in the commensurable {\it and} in the incommensurable
sectors which cannot therefore be associated with the $sl_2$-loop algebra
indicating that the model possesses a   symmetry in the full Hilbert space
 which in the
commensurable sector will  reduce to the $sl_2$-loop algebra.

The BA equations (\ref{ba}) make the implicit assumption,
that none of the factors $S(k_j,k_l)$ becomes
singular or 1. The modulus of some of the $S(k_j,k_l)$
may deviate from 1, corresponding to complex momenta, but
none  is allowed to vanish as long as  $L$ is finite.

 This is not the case on  the {\it infinite} line, where an S-matrix
can vanish (or diverge) signifying a bound state. Consider
  eigenstates of the hamiltonian of the form ($n_1<n_2$):
\be
f(n_1,n_2)=  A_{12}e^{i(p-i\xi)n_1}e^{i(p+i\xi)n_2} +A_{21}
e^{i(p+i\xi)n_1}e^{i(p-i\xi)n_2}.
\label{2pbs}
\ee

Here  $\xi=\log(\cos p/\D)$ so that  $S(p-i\xi,p+i\xi)=0$
rendering this wavefunction  normalizable --  $A_{21}=0$.
Equivalently, $S(p+i\xi,p-i\xi)$
diverges. This state is  therefore a bound state of two ``magnons''
above the ferromagnetic reference state $|0\rangle$.
There exist in general bound states with an arbitrary number $N$ of magnons
in the infinite system, parameterized by complex momenta $k_1,\ldots,k_N$
and the property that
\be
S(k_j,k_{j+1})=0,\ \ \  j=1,\ldots,N-1.
\label{bscon}
\ee

In the finite system, these states are replaced by the so-called
string solutions of (\ref{ba}).
The momenta $k_j$ belonging to
a string do not satisfy (\ref{bscon}), (as all $S$-matrices
must be nonzero for finite $L$) but may approach zero like $e^{-L}$,
as $L$ goes to infinity for fixed $M$.

One would conclude that
for finite $L$ the singular case $S(k_j,k_l)=0$ can never happen for any
two of the parameters in the Bethe wavefunction (\ref{wavef}).
\footnote{Indeed, for a state with only two down spins like (\ref{2pbs}), the
amplitude $A_{21}$ can not be zero on a finite
ring, because in this case the region $n_1<n_2$
can not be distinguished from $n_2<n_1$. If $S(k_1,k_2)$
vanishes, $S(k_2,k_1)$ has to be zero as well, which is impossible
because $S(k_1,k_2)=S(k_2,k_1)^{-1}$.
This case corresponds to $N=2$ ($\D =0$) treated in \cite{mcc1}.
The ``pairs'' of spin-down excitations are of course not bound states
as the interaction is zero.}
We will show, however, that for $q^{2N}=1$ and
$N\ge 3$ there exist  excitations composed of
$N$ spin down magnons satisfying,
\be
S(k_j,k_{j+1})=0,\ \ \ j=1,\ldots,N \ \ \ k_{N+1}=k_1.
\label{cbscon}
\ee
Such states we shall call {\it cyclic bound states} (cbs), because
of the similarity between (\ref{bscon}) and (\ref{cbscon}). In contrast
to the ordinary bound states with property (\ref{bscon}), they exist
on a {\it finite} ring of length $L$. The reason is, that all the
$N$ ``particles'' making up the bound state can not penetrate
each other ($S(k_j,k_{j+1})=0$) so that the amplitudes $A_P$ are nonzero
only for the $N$ cyclic permutations of $1,2,\ldots ,N$.
These states can not be
obtained from the Bethe equations (\ref{ba}), because the
equation determining a momentum $k_j$ belonging to this state
would contain the factor $S(k_j,k_{j+1})S(k_j,k_{j-1})$, i.e.
the product of zero and infinity, rendering it meaningless.
These states have the Bethe ansatz form (\ref{wavef}), but their
parameters are {\it not} given by a solution to (\ref{ba}).
\footnote{They were first found by Baxter when considering
 the Q-T functional equations in \cite{bax}.
 In \cite{mcc2} they were termed ``exact complete
$N$-strings''. This terminology is somewhat misleading
because a string is by definition a
certain solution to the BA equations (\ref{ba}), which is not the case here. }

The physical interpretation of the
corresponding eigenvectors is quite clear: they are
a special type of bound states existing in a finite system,  in contrast to
the usual situation.

We see that the Bethe ansatz
is, strictly speaking, more general than the
Bethe ansatz equations (\ref{ba}): It is impossible to interpret
the cbs as some ``singular'' type of solution to (\ref{ba})
as in the $SU(2)$ or $U_q(sl_2)$ case,
because it will turn out that it contains a free parameter, not
determined by any equation.

We  demonstrate now, that for a cbs with $N$ members to exist
we must have $q^{2N}=1$, or, in the parameterization (\ref{smal}),
\be
\g=m_\g\frac{\pi}{N},\ \ \ 1\le m_\g\le N-1.
\ee
Using (\ref{sma}) and notation $x_j=e^{ik_j}$,
the cbs-condition (\ref{cbscon}) reads
\bea
x_1+x_2^{-1}&=&2\D\nn\\
x_2+x_3^{-1}&=&2\D\nn\\
&\cdots&
\label{cbscon2}\\
x_N+x_1^{-1}&=&2\D\nn
\eea
Using the alternative representation (\ref{smal}),
we get $\Re(\l_j-\l_l)=0$, for all $j,l$, which means the real parts
of the $\l_j$ coincide. For the imaginary parts we have
\bea
\Im(\l_1-\l_2)=2+n_1\frac{2\pi}{\g}\nn\\
\Im(\l_2-\l_3)=2+n_2\frac{2\pi}{\g}\nn\\
&\cdots&
\label{cbscon3}\\
\Im(\l_N-\l_1)=2+n_N\frac{2\pi}{\g}\nn
\eea
This entails
\be
0=2N+\frac{2\pi}{\g}\sum_{i=j}^Nn_j \ \ \ {\textrm{or}}\ \ \
 \g=-\frac{\sum_{j=1}^Nn_j}{N}\pi.
\ee
Because $0<\g<\pi$, we have $1\le -\sum n_j=m_\g\le N-1$.
Using the non-uniqueness of the parameterization
($x_j=x(\l_j)=x(\l_j+2\pi in/\g)$ for $n$ integer), we set the
$n_j=0$ for $j=1,\ldots N-1$ and conclude
$\l_1-\l_N=2i(N-1)$. Also note that the string we thus derived is exact -
unlike the conventional one which involves exponential
corrections \cite{brand}.

Usually one  would write now for the N-string,
\be
l_j=\l_0+(N+1-2j)i\ \ \ j=1,\ldots,N
\ee
 yielding a string symmetric w.r.t the real axis.
However, in our case the string can be shifted along the
imaginary axis by an arbitrary value:
\be
l_j=\l_0+ci+(N+1-2j)i\ \ \ j=1,\ldots,N
\label{string1}
\ee
This incorporates the ``odd parity string'' \cite{taka},
but $c$ is not restricted
to the value $\pi/\g$.

This cbs-string is a transparent excitation, since
from (\ref{string1}) and (\ref{smal}) it follows that:
\be
S({\textrm{cbs}},\L)=\prod_{j=1}^NS(\l_0+ic+(N+1-2j)i,\L)=1
\label{transp}
\ee
for arbitrary $\L$.

The total momentum $P$ satisfies,
\be
e^{iP}=\prod_jx(\l_0+ic+(N+1-2j)i),
\ee
so that
\be
P =n\pi,\ \ \ {\textrm{with}}\ \ \ n \equiv (m_\g+N)\  \mod 2.
\ee
The total excitation energy for the cbs reads
\be
E_{cbs}=2\left(\sum_{j=1}^{N}\cos k^{cbs}_j-\D\right) =0.
\ee
where the last equality follows from eqn (\ref{cbscon2}).

How many independent cyclic bound states are there?
We note that the cbs - parameter $\l_0+ic$
(or alternatively one of the $k^{cbs}_j$, say $k^{cbs}_1$) is {\it not} fixed
by the periodic boundary conditions.
The presence of other particles with arbitrary parameters $k_l$
does not determine $k^{cbs}_1$, because of transparency, eqn (\ref{transp}).
It can therefore only be determined through the length $L$ of the
system.
Let us write the wavefunction of the cbs above the reference state,
without other particles:
\be
f(n_1,\dots,n_N)=A_1x_1^{n_1}x_2^{n_2}\ldots x_N^{n_N}
+ A_2x_2^{n_1}x_3^{n_2}\ldots x_1^{n_N}
+\ldots A_Nx_N^{n_1}x_1^{n_2}\ldots x_{N-1}^{n_N}.
\ee
Periodic boundary conditions lead to,
\bea
A_N&=&A_1x_N^L\nn\\
A_1&=&A_2x_1^L\nn\\
   &\cdots&
\label{period}\\
A_{N-1}&=&A_Nx_{N-1}^L\nn
\eea
hence,
\be
1=\left(\prod_j^Nx_j\right)^L\ \ \ {\textrm{or}}\ \ \ PL\equiv 0\  \mod 2\pi.
\ee
Hence for $P=\pi$, the cbs exist on chains with even length,
and for $P=0$, $L$ is arbitrary.
This remains the only restriction  from periodicity.
The cbs parameter $x_1=e^{ik^{cbs}_1}$ is not fixed by any
constraint and can be chosen as arbitrary
complex number. This result is
quite unusual, as one would expect all parameters of
a Bethe state to be uniquely determined, apart from states
corresponding to ``roots at infinity''. Indeed, in
\cite{mcc2,mcc3}, the attempt is made to find additional
conditions  determining the cbs parameter. We see now, that
this not necessary. To the contrary: It is exactly the
freedom to choose this parameter at will, which allows to find an
analytical formula for the degeneracies in the spectrum
caused by the presence of the cbs.

For each state $|\Psi\rangle$ given by a solution to (\ref{ba}), i.e.
without cbs, there is in principle a continuous set of other
states, having one or more additional cbs with
parameters $x_1, x_1', x_1'', \ldots$. Only
a finite number of them are linearly independent.
The  problem is then to determine
the dimension of the ``one-particle'' Hilbert space, the number
of linear independent cbs above the ``parent'' state
$|\Psi\rangle$, i.e.
in the sector with $S^z=S^z(\Psi)-N$.
Because each cbs is transparent, it can be ``added'' to an arbitrary
eigenstate $|\Psi(\{k^0_l\})\rangle$ of the
hamiltonian without changing the BA equations
for the parameters $\{k^0_l\}$
of $|\Psi\rangle$.
In this way a new state
$|\Psi'(\{k^0_l,k_j\})\rangle$ is created, which has the
same energy as $|\Psi\rangle$ but different spin: $S^z(\Psi')=S^z(\Psi)-N$.
Its rapidities are the cbs parameters $k_1,\ldots,k_N$ and the
rapidities $k^0_1,\ldots,k^0_{L/2-S^z(\Psi)}$ of $|\Psi\rangle$.
However, the amplitudes $A_P(\Psi')$ deviate in a complicated
way from $A_P(\Psi)$, because the $S$-matrices between each member of the
cbs and the  excitations in $|\Psi(\{k^0_l\})\rangle$, $S(k_j,k^0_l)$,
depend now not only on the set $\{k_j\}$ of the cbs but on
$\{k^0_l\}$ as well. This is in contrast to the states generated by
$U_q(sl_2)$, where these $S$-matrices are constant. The cbs
apparently cannot be generated by an operator like (\ref{sl2loopgen}).
The coproduct structure of the quantum group allows only
excitations having $S$-matrices which are independent of the spectral
parameter. The same remains true for the affinization
of $U_q(sl_2)$, $U_q(A_1^{(1)})$, but this question
has to be further investigated.

The dimension of the ``one-particle'' cbs-multiplet built over
a parent state $|\Psi\rangle$ (without any cbs) depends in general
on the spin $S^z(\Psi)$ of the parent state $|\Psi\rangle$,
as well as on the nature of parameters $k^0_1,\ldots,k^0_{L/2-S^z(\Psi)}$
characterizing it, more specifically, whether or not exceptional momenta are
present. We shall find that the parent states fall into three classes.

In the commensurable sectors the parent state can not contain
the exceptional values $\pm k_q$ among the momenta $\{k^0_l\}$.
However in an incommensurable sector with $S^z(\Psi)\equiv (N-m)\mod N$,
$m=1,\ldots N-1$,
the state $|\Psi\rangle$ may contain $m$ momenta which are either
all $k_q$ or all $-k_q$.\footnote{
This possibility was pointed out to us by
B.McCoy; these momenta manifest themselves as
``roots at infinity'' in the parameterization
(\ref{l-param}).}
Because the $S$-matrix for these excitations takes
then the constant value (\ref{smatrixq}), the number of
independent cbs that can be added to $|\Psi\rangle$ is modified
with respect to the case where the set $\{k^0_l\}$ contains
no ``exceptional'' momenta $\pm k_q$.

Another way of having  exceptional momenta in the parent state
is in the form of  $m$ pairs of exceptional momenta
with opposite sign, $\{k_q,-k_q\}$, added to the state $|\Psi_0\rangle$
to create a state $|\Psi\rangle$, degenerate with $|\Psi_0\rangle$.
The BA equations for $|\Psi\rangle$ read,
\bea
e^{-ik^0_lL}&=&\prod_{l'\neq l}^{M'}S(k^0_l,k^0_{l'})(q^{-2}q^{2})^m\nn\\
q^{-L}&=&(q^{-2})^{M'+m}\nn\\
q^{L}&=&(q^{2})^{M'+m}.\nn
\eea
We have $S^z(\Psi_0)=L/2-M'$ and $S^z(\Psi)=L/2-M'-2m$
from which follows,
\be
S^z(\Psi_0)\equiv m\ \mod N \ \ \textrm{and}\ \ \ S^z(\Psi)\equiv -m\ \mod N.
\label{pairsq}
\ee
This type of degeneracy can be reduced to the case where  cbs excitations are
present by using
the $Z_2$-invariance of the spectrum with respect to
flipping the $z$-component of all spins, $S^z\ra -S^z$. Whereas
the cbs-multiplets in commensurable sectors are mapped
onto themselves by this
transformation, there are two different multiplets in the
incommensurable case, one corresponding to
$S^z\equiv m\ \mod N$, the other to $S^z\equiv -m\ \mod N$.
All states in the two sets are energetically degenerate.
The states of the first set are generated from the reference
state with all spins up and the members of the second  are
the spin-flipped states. From (\ref{pairsq}) we see that if
$|\Psi_0\rangle$ belongs to the first set, $|\Psi\rangle$
belongs to the second and can be described as generated
in the spin-flipped representation by adding one or more
cbs to the state $|\Psi_0'\rangle$ with $S^z(\Psi_0')=-S^z(\Psi_0)$.
If we fix the representation (by using the spin-up reference state),
the state $|\Psi\rangle$ can be regarded as parent state
(because it is not a member of the multiplet generated from
$|\Psi_0\rangle$) and adding one cbs to $|\Psi\rangle$
results again in a modification of the dimension
formula as above.

We have therefore three different classes of degeneracies,
depending on the presence of exceptional momenta in
the parent state:\\
I)\ \ The parent state contains no exceptional momenta.
This is always the case for commensurable sectors.\\
II)\ \ The parent state has $S^z(\Psi)\equiv m\ \mod N$
($m=1,\ldots N-1$) and contains either $N-m$ equal momenta
$ k_q$ or $N-m$ equal momenta $-k_q$.\\
III)\ \ The parent state has $S^z(\Psi)\equiv m\ \mod N$
($m=1,\ldots N-1$) and contains $N-m$ momenta $k_q$ and $N-m$ momenta $-k_q$.

These three cases are the
reason for the three types of degeneracies observed in
the incommensurable sectors \cite{mcc1}. The case with
independent numbers of $k_q$ and $-k_q$ can not occur as is
easily seen from the BA equations.

\bigskip

In the following we will concentrate on the case $N=3$.
We have $\D=\pm 1/2$, corresponding to $\g=2\pi/3, \pi/3$.
For $\D=1/2$, the cbs exist only on chains of even length $L$ ($P=\pi$)
and for $\D=-1/2$ on all chains ($P=0$).

We shall first determine the dimension dim $\H_1^0$ of the
single-cbs space above the reference state $|0\rangle$ with all spins up, and
subsequently  the dimension of the single-cbs
space $\dim\H^{\Psi}_1$ built over a general state  $|\Psi\rangle$.
The amusing
combinatorics is presented in appendix B.

Our results are  the following:
The dimensionality of the cbs-space over $|0\rangle$ is,
\be
\dim\H^0_1= \m(L-2)+\iota(L-2),
\label{mult}
\ee
where for an integer $n$ we define the integers $\m(n)\in\{0,1,2,3,\ldots\}$
and $\iota(n)\in\{0,1,-1\}$ through the relation,
\be
n=3\m(n)+ \iota(n).
\label{index}
\ee

The dimensionality of the degenerate space
above a  state $|\Psi\rangle$  -- different from the reference state --
falls into the three cases discussed above,\\
Case I): \be \dim\H^{\Psi}_1=\m(2S^z(\Psi)-2)+\i(2S^z(\Psi)-2).
\label{mult2} \ee Case II): \be
\dim\H^{\Psi}_1=\m(2S^z(\Psi)-2)+\i(2S^z(\Psi)-2)+1. \label{mult3}
\ee Case III): \be
\dim\H^{\Psi}_1=\m(2S^z(\Psi)-2)+\i(2S^z(\Psi)-2)+2. \label{mult4}
\ee Note that the dimension of $\dim\H^{\Psi}_1$ in cases II) and
III) do not depend on $m$, (which for $N=3$ may take the values 1
and 2.) The derivation of (\ref{mult2}-\ref{mult4}) is also given
in appendix B.

Consider now states with two or more cbs. 
These correspond to multiplicities of type (\ref{degNdiv}), (\ref{degNndiv})
with $l>1$. 
Whereas it is possible in principle
to determine the number  of linear
independent states containing $l$ cbs over a state $|\Psi\rangle$ along the
same lines as done for one cbs in appendix B, there are many more
entangled terms in the wavefunction  and the
calculation is very cumbersome. Because a mapping between the
cbs and the states generated by $sl(2)$-loop is not known at present,
we are unable to make contact with the program outlined in
\cite{mcc2}, which attempts to use representation theory of this
algebra to compute the dimensions of the multiplets in
commensurable sectors.
 
Nevertheless, we wish to present the following argument,
which renders formula (\ref{mult5}) below at least plausible. 
Let us introduce the notion of
the ``effective size'' of a particle and  begin by illustrating
its usefulness in computing the dimension of a degenerate space
by applying it to the simple case of an XXX model. The dimension
of the degenerate spaces of the XXX model is obtained by
assuming that the ``effective size'' of each ordinary particle
(spinless fermion/spin-down excitation) is two lattice sites
while the ``effective size'' of a transparent excitation is not
two but one lattice site, as follows from the fact that the
number of allowed $k=0$ excitations which can be added to a state
with $M$ particles is then given by $L-2M$.  We may add
transparent excitations until the lattice is ``completely
filled''. In other words, the ``effective length'' of the ring
available for transparent excitations is reduced by an ordinary
particle by two and by a transparent particle by one. Because
$L-2M=2S_{max}^z$, we find for the dimension of the
$SU(2)$-multiplet, $2S^z_{max}+1$ (including the parent state),
which is correct, because the parent state is highest weight for
$SU(2)$. That such a counting works is of course due to a
non trivial property of the parent state, namely to be highest
weight for the symmetry group $SU(2)$.

Lets assume now, that the same argument applies to the cbs symmetry. We have
seen that the ``effective length'' of the ring entering formulas
(\ref{mult2}) - (\ref{mult4}) for the one-cbs space is again
$L-2M$. By analogy to the $SU(2)$ case we would conclude that the
effective size of ordinary particles is 2 and the effective size
for the cbs  is 3  - the minimal size for a three-particle
excitation.  Consider then a parent state $|\Psi\rangle$ with spin
$S^z$ and one additional cbs. To calculate the dimension for a
second cbs, we have to take into account the reduction of the
effective ring length $L_{eff}=L-2M=2S^z$ caused by the first cbs:
$L_{eff}'=L_{eff}-3$. The dimension formula for the second cbs
would read,
 \be
 \m(2S^z-3-2)+\i(2S^z-3-2)=\dim\H^{\Psi}_1-1.
 \ee
Because both cbs are indiscernible we have,
 \be
\dim\H^{\Psi}_2=\frac{\dim\H^\Psi_1(\dim\H^\Psi_1-1)}{2}.
 \ee
Repeating this argument we conclude that if the
one-cbs space has dimension $\dim\H^\Psi_1$, the number of
available states for $l$
cbs above the state $|\Psi\rangle$ reads:
\be
\dim\H_l^\Psi={\dim\H^\Psi_1 \choose l}. \label{mult5}
\ee
The  dimension of a complete multiplet in the commensurable sector, 
$2^{l_{max}}$ with $l_{max}=\dim\H^\Psi_1$, follows then  from
(\ref{mult5}) and formulae (\ref{mult2}-\ref{mult4}) for the
single-particle space.

This argument does not constitute a proof of the binomial
formula (\ref{mult5}), as this would require in addition the
demonstration that $|\Psi\rangle$ is
highest weight for a (unknown) symmetry algebra
different from $sl(2)$-loop, as our parent states are clearly
{\it not} highest weight for this algebra \cite{mcc2}.  

We wanted to draw attention to the curious fact, that such a
simple counting indeed leads to the numerically observed multiplicities,
not only for simple groups (as $SU(2)$), but 
even in the present case, where the underlying symmetry
is much more complicated.

Let's return to the case with a single cbs and check eqn
(\ref{mult}). For $L\equiv 0\  \mod 3$, we have \be
\m(L-2)=\frac{L}{3}-1, \ \ \ \iota(L-2)=1, \ee therefore, in the
commensurable case, \be \dim\H^0_1=\frac{L}{3}, \ee in accord
with (\ref{degNdiv}). We give in Table 1) the number of states in
the sector $S^z=L/2-3$, which are degenerate with the reference
state, for $L$ between 3 and 18 and $\D=-1/2$:\par \vspace{4mm}
\noindent Table 1):\par \vspace{3mm} \noindent
\begin{tabular}{c|c|c|c|c|c|c|c|c|c|c|c|c|c|c|c|c}
$L$&\ 3\ &\ 4\ &\ 5\ &\ 6\ &\ 7\ &\ 8\ &\ 9\ &\ 10\ &\ 11\
&\ 12\ &\ 13\ &\ 14\ &\ 15\ &\ 16\ &\ 17\ &\ 18\ \\
\hline
deg($L/2-3$)&1&0&1&2&1&2&3&2&3&4&3&4&5&4&5&6
\end{tabular}\par
\vspace{4mm}
\noindent
These (numerically confirmed) degeneracies coincide with the
values predicted by (\ref{mult}) in the commensurable and
incommensurable cases. Note the non-monotonic growth of the dimension
of $\H^0_1$: e.g. for $L=6$ there are two independent cbs, whereas for
$L=7$ there is only one independent state.
As the reference state contains zero excitations, only case I) above
is relevant for commensurable and incommensurable chain lengths.

\section{Conclusions}

We have shown that the XXZ spin chain at roots of the unity admits a class of
unusual excitations, the cyclic bound states. These excitations are transparent
to other particles in the sense defined above and lead to a rich degeneracy
pattern associated with some quantum symmetry.
In particular we derived analytic expressions for
the dimensionality of the degenerate spaces for $N=3$ and $l=1$.

The presence of transparent solutions indicates  underlying symmetries.
In particular the $sl_2$ - loop symmetry contains generators
$S^{-(N)},T^{-(N)}$
which are associated with the
 special solution
of (\ref{cbscon2})
having  all the $x_1\ldots x_N$ equal: $x_1=x_2=\ldots x_N=q^{\pm 1}$.
Applying $S^{-(N)},T^{-(N)}$ to a state $|\Psi\rangle$ means
adding $N$ exceptional momenta $\pm k_q$.
For these states both numerator and denominator of
$S(k_j,k_{j+1})$ vanish, which means that $S(k_j,k_{j+1})$ has
to be 1, not zero.
This is the reason these states exist only in the commensurable sectors.
However, they are linearly dependent on the general cbs:
 The states generated
by $sl_2$-loop in the commensurable sectors are ``singular'' cyclic
bound states. If they span the complete multiplet in the
commensurable sectors, it must be possible to write the
general cbs as  superposition of the special states.
As the explicit form of the states generated by $sl(2)$-loop is not
known in general (see \cite{mcc1,mcc2,mcc3}) this question can not
be answered at present.

Section C of the Appendix contains
some  examples for
the multiplets in the commensurable and incommensurable cases
for chain lengths between 8 and 14. It is apparent, that
formulas (\ref{mult2}-\ref{mult4}) describe only part
of the degeneracies, namely that part, which is
related to the cyclic bound states. First, we note that
some of the energies are two-fold degenerate already
in the ``parent sector'', f.e. in Table 3 the energies
-0.4142, 1.0 and 2.4142. This degeneracy is due to parity
invariance, the two degenerate states have opposite
momentum $P$. Accordingly, a factor of 2 multiplies the
dimensions obtained from (\ref{mult2}) in the sector
$S^z=0$. 

This parity doubling occurs as well for several
energies in Tables 4 -- 10. Some energies show even
a higher degeneracy in the parent sector: Energy 0.0 in
Table 4, energy 0.5 in Table 6 and energy 1.0 in Table 8.
This degeneracy of the parent states has to be taken into
account, if one compares with formulae (\ref{mult2}-\ref{mult4}).
Example: The energy 0.5 in Table 6 is three-fold degenerate
in the parent sector $S^z=3$, formula (\ref{mult2}) gives
a degeneracy of 2 for each state in sector $S^z=0$, which yields
a total dimension of 6 for this energy in sector $S^z=0$.

However, not {\it all} degeneracies can be explained by using
the degeneracy of the parent states. F.e. energy 1.0 in
Table 8 should be 10--fold degenerate in sector $S^z=1$ and
5-fold degenerate in sector $S^z=-2$. Instead we find a
12-fold degeneracy for $S^z=1$  and 9-fold for $S^z=-2$.  
These additional degenerate states are not due to the
presence of cbs but have a different origin (which we do
not know).

In any case the predicted degeneracies from the cbs
give a lower bound to the number of energetically degenerate states.

Tables 7 and 8 give  examples for case II): There are two states with energy
3.0 in the sector $S_{max}^z=5$, corresponding to one-particle
excitations with $\pm k_q$. They exist because $q^{12}=1$.
Formula (\ref{mult3}) yields the degeneracy $2+1=3$ for each state
in the sector $S^z=5-3=2$ and (\ref{mult5}) gives ${3 \choose 2}=3$ in
the sector $S^z=5-6=-1$, which coincides with the numerical results
in table 7. 
$S_{max}^z=4$ corresponds to two excitations in the
parent state and we have again two states with energy 3.0. These are
the states with $(k_q,k_q)$, resp. $(-k_q,-k_q)$. The degeneracy in the
sector with one cbs ($S^z=1$) is again 3 and therefore the same
in the sector with two cbs ($S^z=-2$). This is shown in table 8.

Table 4 contains a state with a pair of exceptional momenta
with opposite sign (case III),
the state with energy 2.0 in sector $S^z_{max}=2$.
The generic degeneracy in the sector $S^z=-1$
according to formula (\ref{mult2}) is
zero but the state with energy 2.0 possesses two states with
$S^z=-1$, having the same energy, because eqn (\ref{mult4})
yields the value $0+2=2$ for the dimension of $\H^{\Psi}_1$.

\section{Acknowledgments}

Part of the work was carried out while N. A. was visiting
 the theory group at ENS - Lyon. He  wishes to thank the members of the group
for their kind hospitality, and to J.-M. Maillet and L. Freidel
for many enlightening discussions. We are most
grateful to B. McCoy for insightful
comments on an early version of this manuscript.

\section{Appendix}
\subsection{Example:  Adding transparent particles in the XXX-Chain.}

Take $l=2$ and $M=3$, then take some $Q$ from $S(5)$, say:
\[
Q=\pmatrix{1&2&3&4&5\cr 4&3&2&5&1}
\]
We want to construct a permutation $P$, which ``coincides''
with $Q$ regarding the first three momenta $k_1, k_2, k_3$, i.e.
$n(k_3)<n(k_2)<n(k_1)$. This is the permutation
\[
P(Q)=\pmatrix{1&2&3\cr 3&2&1}
\]
Now we apply formula (\ref{permp}). Compute first $P_{-1}(1)$:
$Q^{-1}(1)=5$. Now look for the preimages of 2 and 3 under $Q$:
$Q^{-1}(2)=3$ and $Q^{-1}(3)=2$. Both are less than 5 (the
preimage of 1) and therefore the number we are looking for is 2+1=3.
(Of course, the 1 itself with $Q^{-1}(1)=5$ is counted as well.)
It follows that $P^{-1}(1)=3$.\\
Similar: because $Q^{-1}(2)=3$, there are two numbers $k$, less or equal to 3,
for which $Q^{-1}(k)\le Q^{-1}(2)$, namely 3,
(with $Q^{-1}(3)=2<3$) and 2 itself. It follows $P^{-1}(2)=2$.\\
Again: As $Q^{-1}(3)=2$, there is only one index $k$ which satisfies the
condition $Q^{-1}(k)\le Q^{-1}(3)=2$, namely 3 itself. Therefore
$P^{-1}(3)=1$ and we get the wanted permutation $P(Q)$.

\subsection{Derivation of eqn (\ref{mult})
and (\ref{mult2} -- \ref{mult4}).}
We confine ourselves here to the case $\D=1/2$, the case
$\D=-1/2$ being completely analogous.

We begin by determining the parameters $x_2,x_3$ of the
cbs in terms of $x_1=z$. From (\ref{cbscon2}) we get
\be
x_2=(1-z)^{-1} \ \ \ x_3=1-z^{-1}.
\ee
To construct $\H^0_1$
we calculate the wavefunction of the cbs,
\be
f(n_1,n_2,n_3;z)=A_1x_1^{n_1}x_2^{n_2}x_3^{n_3}
+A_2x_2^{n_1}x_3^{n_2}x_1^{n_3}
+A_3x_3^{n_1}x_1^{n_2}x_2^{n_3}.
\label{wavefa}
\ee
We have (\ref{period}),
\be
A_1=1\ \ \ \ A_2=x_1^{-L}\ \ \ \ A_3=x_3^L.
\ee
It follows
\bea
f(n_1,n_2,n_3;z)&=&(-1)^{n_3}z^{n_1-n_3}(1-z)^{n_3-n_2}\nn\\
  &+&(-1)^{n_2}z^{n_3-n_2-L}(1-z)^{n_2-n_1}\label{f1}\\
&+&(-1)^{L+n_1}z^{n_2-n_1-L}(1-z)^{L+n_1-n_3}.\nn
\eea
We see, that $f$ is a certain meromorphic function in $z$,
parameterized by the set of integers $\{n_1,n_2,n_3\}$ with
$0\le n_1<n_2<n_3\le L-1$.\\
Now, as the parameter $z$ is arbitrary, the question how many
of the vectors $|\ps(z)\rangle=f(n_1,n_2,n_3;z)|n_1,n_2,n_3\rangle$
are linear independent, is equivalent to ask, how many of the
functions $f(n_1,n_2,n_3;z)$, indexed by the set $\{n_1,n_2,n_3\}$,
are linear independent over $\CC$. (This follows from the
equality of column rank and row rank of a matrix.)\\
We start out with a total of ${{L \choose 3}}$
different functions $f(n_1,n_2,n_3;z)$. This is the maximal number
of possible linear independent cbs states in the given sector of the Hilbert
space. However, the set of really independent functions of
type $f(n_1,n_2,n_3;z)$ is much smaller. First, we have the
translation property, following from $P=\pi$. As one sees from (\ref{f1}),
we have
\be
f(n_1,n_2,n_3;z)=-f(n_1+1,n_2+1,n_3+1;z).
\ee
Therefore, we can fix $n_1=0$. Furthermore, we put
\bea
n_2-1&=&i\nn\\
n_3-n_2-1&=&j\nn\\
L-n_3-1&=&k\nn
\eea
we have $i+j+k=L-3$ and $0\le i,j,k\le L-3$. Then
\be
f(i,j,k;z)=\frac{z(z-1)}{z^L}\p(i,j,k;z)
\ee
with
\be
\p(i,j,k;z)=(z-1)^iz^j-(-1)^i(z-1)^jz^k+
(-1)^j(z-1)^k(-z)^i.
\label{f2}
\ee
Now one confirms that
\be
\p(i+1,j,k;z)=\p(i,j+1,k;z)-\p(i,j,k+1;z).
\label{relphi1}
\ee
We can, without loss of generality, put $i=0$.
Lets look at the polynomials
\be
\varphi_j(z)=\p(0,j,L-3-j;z)=z^j-(z-1)^jz^{L-3-j}-
(1-z)^{L-3-j},
\label{f3}
\ee
$j=0,\ldots L-3$, the sign of the last summand is negative
because $L$ is even.
The maximal number of independent states is reduced in this way
to $L-2$. However, the set $\{\varphi_j(z)\}_{j=0,\ldots L-3}$ is not linear
independent. There are further relations among the polynomials.
Lets form the expression
\be
\P(x,z)=\sum_{j=0}^{L-3}x^j\varphi_j(z){ L-3 \choose j}.
\label{f4}
\ee
This is a way to shift the dependence on $j$ over to the
(complex) parameter $x$. We find:
\be
\P(x,z)=(xz+1)^{L-3}-((x+1)z-x)^{L-3}-(x+1-z)^{L-3}.
\label{f5}
\ee
What relations are possible among the functions $\P(x,z)$?
We make the ansatz,
\be
\P(x,z)=\sum_r\a_r\P(\x_r,z).
\label{rel1a}
\ee
By examination of (\ref{f5}) and (\ref{rel1a})
one sees that all possible relations of this type reduce to the
following:
\be
\P(x,z)=\a^{L-3}\P(\x,z).
\label{rel1}
\ee
To solve (\ref{rel1}) identical in $z$, we must have
\bea
xz+1&=&-\a((\x+1)z-\x)\nn\\
(x+1)z-x&=&\a(\x+1-z) \label{rel2}\\
x+1-z&=&-\a(\x z+1),\nn
\eea
and
\be
\x=-\frac{1}{x+1},\ \ \ \ \a=-(1+x).
\label{rel3}
\ee
Then
\be
\P(x,z)=-(1+x)^{L-3}\P(-(1+x)^{-1},z).
\label{rel4}
\ee
Going now back to (\ref{f4}), we can write
(\ref{rel4})
as
\be
\sum_{j=0}^{L-3}c_j(x){\tilde\p_j}(z)=0,
\label{rel5}
\ee
with
\be
c_j(x)=x^j-\frac{(-1)^{j+1}}{(1+x)^j}(1+x)^{L-3},\ \ \
{\tilde\varphi_j}(z)={L-3 \choose j}\varphi_j(z),
\ee
for arbitrary $x$. Because  relation (\ref{rel5})
is valid for all $x$,
we conclude that
\be
\dim\langle\{\varphi_j(z)\}_{j=0\ldots L-3}\rangle
+\dim\langle\{c_j(x)\}_{j=0\ldots L-3}\rangle\le L-2,
\label{dimensions1}
\ee
$\langle\ldots\rangle$ denotes the linear span.
But because {\it all} linear relations among the $\P(x,z)$
can be reduced to (\ref{rel1}), we have actually
\be
\dim\langle\{\varphi_j(z)\}\rangle
+\dim\langle\{c_j(x)\}\rangle= L-2.
\label{dimensions}
\ee
By choosing the basis $x^0,x^1,\ldots x^{L-3}$ to span
$\langle\{c_j(x)\}\rangle$,
we find that its dimension
is equal to the rank of the $(L-2)\times(L-2)$-matrix
\be
\id-A_{L-3},
\label{matr}
\ee
with
\be
A_{n}(i,k)=(-1)^{i+n}{n-i\choose k},
\label{pascal}
\ee
for $i,k=0,\ldots,n$.
The $(n+1)\times(n+1)$-matrices $A_n$, we call
Pascal matrices, for obvious reasons. (We have defined
${n\choose m}=0$ if $ m>n$)
The matrix $A_3$, f.e. reads
\be
A_3=\pmatrix{-1&-3&-3&-1\cr
                 1&2&1&0\cr
                 -1&-1&0&0\cr
                 1&0&0&0}.
\ee
The Pascal matrices have interesting properties,
intimately connected with our problem. Lets define
the inversion Matrix $S_n$ through
\be
S_n(i,k)=\d_{i,n-k}
\ee
for $i,k=0,\ldots,n$.
Then we can show by induction
\be
S_nA_nS_n=A_n^{-1}=A_n^2,
\label{pasrel1}
\ee
from which we have $A_n^3=\id$.
From this follows, that $A_n$ is diagonalizable. Moreover
we have for the trace of $A_n$:
\be
{\textrm{tr}}A_n=\i(n+1)
\label{pasrel2}
\ee
(compare (\ref{index})).
These identities for the Pascal matrices are equivalent to
the combinatorial identities
\be
\sum_{k=0}^n(-1)^k{n-i\choose k}{n-k \choose l}={i\choose n-l},
\ee
\be
\sum_{k=0}^n(-1)^{i+k+n}{n-i\choose k}{k\choose n-l}=\d_{il},
\ee
and
\be
(-1)^n\sum_{k=0}^n(-1)^k{n-k\choose k}=\i(n+1).
\ee
Now we can compute the
characteristic polynomial of $A_n$ for all $n=0,1,\ldots$.
(This is relevant for $\D=1/2$ only for odd $n$, but for
$\D=-1/2$, all $n$ are needed.)
We write
\[
(-1)^{n+1}\det(A_n-\l\id)=\det(\l\id-A_n)
\]
and use
\[
\log\ \det(\l\id-A_n)=\tr[\log(\l\id-A_n)].
\]
Expanding the log, we have,
\be
\log(\l\id-A_n)=\log(\l\id)-\sum_{i=1}^\infty\frac{\l^{-i}}{i}A_n^i,
\label{logpas}
\ee
and
with relations (\ref{pasrel1})
\be
\sum_{i=1}^\infty\ldots = \frac{A_n}{\l}+\frac{1}{2}\frac{S_nA_nS_n}{\l^2}
+\frac{1}{3}\frac{\id}{\l^3}+\frac{1}{4}\frac{A_n}{\l^4}+
\frac{1}{5}\frac{S_nA_nS_n}{\l^5}+\frac{1}{6}\frac{\id}{\l^6}+\ldots
\label{sumpas}
\ee
With (\ref{pasrel2}), $\tr(S_nA_nS_n)=\tr(A_n)$ and $\tr\id=n+1$, it follows
\be
\tr\left(\sum_{i=1}^\infty\frac{\l^{-i}}{i}A_n^i\right)=
(n+1)\sum_{i\in I_3}\frac{\l^{-i}}{i}
+ \i(n+1)\sum_{i\notin I_3}\frac{\l^{-i}}{i}.
\label{sumpas2}
\ee
The index set $I_3$ contains all multiples of 3: $I_3=3,6,9,\ldots$.
Using now $n+1=3\m(n+1)+\i(n+1)$, we rewrite the first sum
on the r.h.s of (\ref{sumpas2}) as,
\be
(\m+\i/3)\sum_{j=1}^{\infty}\frac{(\l^{-3})^j}{j}=-(\m+\i/3)\log(1-\l^{-3})
\ee
and the second sum,
\be
\i\left[\sum_{j=1}^{\infty}\frac{\l^{-j}}{j}-(1/3)
\sum_{j=1}^{\infty}\frac{(\l^{-3})^j}{j}\right].
\ee
Adding both terms and inserting into (\ref{logpas})
we find,
\be
\tr(\log(\l\id-A_n))=\m(n+1)[3\log\l+\log(1-\l^{-3})]+
\i(n+1)[\log\l+\log(1-\l^{-1})]
\ee
and finally,
\be
{\textrm{det}}(A_n-\l\id)=(-1)^{n+1}\left[
(\l^3-1)^{\m(n+1)}(\l-1)^{\iota(n+1)}\right].
\label{det}
\ee
From (\ref{det}) we have the dimension of the eigenspace
to the eigenvalue 1: $\m(n+1)+\iota(n+1)$,
(remember that $A_n$ is diagonalizable).
And this entails, in view of (\ref{dimensions}) and (\ref{matr}), that
\be
\dim\langle\{\varphi_j(z)\}\rangle
=\dim\H^0_1 = \m(L-2)+\iota(L-2).
\ee
Now we consider the case $S^z(\Psi)<L/2$. First, assume that only
one excitation is present:
\be
|\Psi\rangle=\sum_{n=0}^{L-1}x_0^n|n\rangle,
\ee
with $x_0^L=1$ and $x_0\neq q^{\pm 1}$ (case I).
The wavefunction with one cbs on top of the single particle state
$|\Psi\rangle$ reads
\be
|\Psi'\rangle=\sum_{0\le n_0<n_1<n_3<L}
f(n_0,n_1,n_2,n_3)|n_0,n_1,n_2,n_3\rangle
\ee
with
\bea
 &\ph{0}&f(n_0,n_1,n_2,n_3)=\nn\\
 &\ph{0}&
 x_0^{n_0}x_1^{n_1}x_2^{n_2}x_3^{n_3}
+S_{01}x_1^{n_0}x_0^{n_1}x_2^{n_2}x_3^{n_3}
+S_{30}x_1^{n_0}x_2^{n_1}x_0^{n_2}x_3^{n_3}
+x_1^{n_0}x_2^{n_1}x_3^{n_2}x_0^{n_3}\nn\\
  &+&
S_{01}x_0^{n_0}x_2^{n_1}x_3^{n_2}x_1^{n_3-L}
+S_{30}x_2^{n_0}x_0^{n_1}x_3^{n_2}x_1^{n_3-L}
+  x_2^{n_0}x_3^{n_1}x_0^{n_2}x_1^{n_3-L}
+S_{01}x_2^{n_0}x_3^{n_1}x_1^{n_2-L}x_0^{n_3}\label{wavef0}\\
  &+&
S_{30}x_0^{n_0}x_3^{n_1+L}x_1^{n_2}x_2^{n_3}
+  x_3^{n_0+L}x_0^{n_1}x_1^{n_2}x_2^{n_3}
+S_{01}x_3^{n_0+L}x_1^{n_1}x_0^{n_2}x_2^{n_3}
+S_{30}x_3^{n_0+L}x_1^{n_1}x_2^{n_2}x_0^{n_3}\nn
\eea
with the $S$-matrices
\be
S_{01}=-\frac{(x_0-1)z+1}{(z-1)x_0+1},\ \ \ S_{30}=\frac{x_0-z}{(z-1)x_0+1}.
\label{snullmatr}
\ee
Lets define
\bea
a&=&n_1-n_0\nn\\
b&=&n_2-n_1\\
c&=&n_3-n_2\label{abcd}\\
d&=&L+n_0-n_3,\nn
\eea
with $1\le a,b,c,d\le L-3, a+b+c+d=L$.
The polynomial $f'(n_0,n_1,n_2,n_3;x_0,z)=z^L[(z-1)x_0+1]f(n_0,n_1,n_2,n_3)$
can be written as
\bea
 &\ph{0}&f'(n_0,n_1,n_2,n_3;x_0,z)=\nn\\
 &\ph{0}&
(-x_0)^{n_0}[(-1)^{b+a}\p(c,L-b-c,b+1;z)+x_0(-1)^d\p(L-b-c,b,c+1;z)]\nn\\
 &+& (-x_0)^{n_1}[(-1)^{c+b}\p(d,L-d-c,c+1;z)+x_0(-1)^a\p(L-d-c,c,d+1;z)]
\label{wavef3}\\
 &+& (-x_0)^{n_2}[(-1)^{d+c}\p(a,L-a-d,d+1;z)+x_0(-1)^b\p(L-a-d,d,a+1;z)]\nn\\
 &+& (-x_0)^{n_3}[(-1)^{a+d}\p(b,L-a-b,a+1;z)+x_0(-1)^c\p(L-a-b,a,b+1;z)]\nn.
\eea
One sees, that $f'$ lies in the span of the functions
$\p(i,j,k;z)$, with $i+j+k=L+1$ and $i=1,\dots L-3$; $j,k = 2,\ldots L-2$,
respectively $j=1,\dots L-3$ and $i,k = 2,\ldots L-2$. The second case
can be reduced to the first with the aid of the relation
\be
\p(i,j,k;z)=(-1)^{i+1}\p(j,k,i;z).
\label{relphi2}
\ee
Moreover, using (\ref{relphi1}) and (\ref{relphi2}), one can show that
\be
\dim\langle\{\p(i,j,k;z)\}|i\ge 1;j,k\ge 2\rangle=
\dim\langle\{\p(i,j,k;z)\}|i,j,k \ge 1\rangle.
\ee
Lets write $\langle\{\p(i,j,k;z)\}|i,j,k \ge 1\rangle =\H'$.
Repeating the arguments above, we have
\be
\dim\H'=\m(L-1)+\i(L-1).
\ee
Now for the dimension of $\H^{\Psi}_1=\langle f'(n_0,n_1,n_2,n_3;x_0,z)\rangle$
we have
\be
\dim\H^{\Psi}_1=\dim\H'-1,
\label{dim2}
\ee
because
the functions of type (\ref{wavef3}) do not span all of $\H'$:
they satisfy an additional relation coming from the fact, that
$x_0^L=1$, i.e. that $|\Psi\rangle$ is an eigenstate of the hamiltonian.
To prove this, assume the contrary, $\H'\subset \H^{\Psi}_1$.
Then one could write
\be
f'(n_0,n_1,n_2,n_3;\x,z_0)
=\sum_k \a_kf'(n_0,n_1,n_2,n_3;x_0,z_k)
\ee
with {\it arbitrary} $\x$ and $z_0$ for some $z_k$ and all
$\{n_0,n_1,n_2,n_3\}$. But this is impossible, because all the
$f'(n_0,n_1,n_2,n_3;x_0,z_k)$ satisfy the periodicity condition
\be
f'(0,n_1,n_2,n_3;x_0,z_k)=f'(n_1,n_2,n_3,L;x_0,z_k)
\ee
whereas this is not true for $f'(n_0,n_1,n_2,n_3;\x,z_0)$ if $\x^L\neq 1$.
This additional relation is a reflection of the fact, that $|\Psi'\rangle$
satisfies exactly one BA equation (\ref{ba}).
Besides this one, there is no other relation among the functions
$f'$ because $x_0$ can now be considered as
a free parameter in (\ref{wavef3}). In an analogous way, $r$ excitations
in $|\Psi\rangle$ , with parameters $x_0,x_1,\ldots x_{r-1}$,
lead  for $\H'$ to
the dimension formula,
\be
\dim\H'=\m(L-2+r)+\i(L-2+r)
\ee
and
\be
\dim\H^{\Psi}_1=\dim\H'-r,
\label{dimr}
\ee
because now  $|\Psi\rangle$ and $|\Psi'\rangle$ satisfy $r$
BA equations.
(\ref{dimr}) can be written in another way, by
observing that
\[
\m(n+3m)-\m(n)=m\ \ \ \ \i(n+3m)=\i(n),
\]
for all $n,m\ge 1$.
It follows
\be
\dim\H^\Psi_1=\m(L-2-2r)+\i(L-2-2r)=\m(2S^z(\Psi)-2)+\i(2S^z(\Psi)-2).
\label{dim3}
\ee
This completes the proof of (\ref{mult2}) corresponding to case I).

For case II) we set $x_0=q$. To have $x_0^L=1$, $L$
must be divisible by 3. Now, instead of (\ref{snullmatr}), we
have
\be
S_{01}=q^{-2},\ \ \ S_{30}=q^2.
\ee
These $S$-matrices do not depend on $z$, which allows to write
for the component of $f(n_0,n_1,n_2,n_3)$ multiplying $x_0^{n_0}$:
\bea
\tilde{f}(a,b,c,d;q,z)=&\ph{q^{-2}(-1)^{a+b}z^c(1-z)^b+}
&\ph{ q^2(-1)^{L+a}z^b(1-z)^{d+a}}\nn\\
z^{-L}[(-1)^{a+b+c}z^{d+a}(1-z)^c
+&q^{-2}(-1)^{a+b}z^c(1-z)^b+& q^2(-1)^{L+a}z^b(1-z)^{d+a}],
\label{wavefq}
\eea
where we have used convention (\ref{abcd}). Cyclic permuted
expressions are obtained for the components of $f$ multiplying
$x_0^{n_i}$, $i=1,2,3$. To compute $\dim \H'$, consideration of
$\tilde{f}$ in (\ref{wavefq}) is sufficient.
We have,
\be
\tilde{f}=(-1)^az^{-L}\p_q(i,j,k;z),
\ee
with
\be
\p_q(i,j,k;z)=q^{-2}(z-1)^iz^j+(-1)^i(z-1)^jz^k + q^2(-1)^j(z-1)^k(-z)^i,
\ee
with $i+j+k=L$; $i,j=1,\ldots L-3$; $k=2,\ldots L-2$.
The polynomials $\p_q(i,j,k;z)$ fulfill relation (\ref{relphi1})
and (\ref{relphi2}) reads now
\be
\p_q(j,k,i;z)=(-1)^iq^{-2}\p_q(i,j,k;z),
\ee
because $q^6=1$.
From these relations we have
\be
\dim\langle\{\p_q(i,j,k;z)|i,j\ge 1;k\ge 2\}\rangle=
\dim\langle\{\p_q(i,j,k;z)|i,j,k\ge 1\}\rangle=\dim \H'.
\ee
$\H'$ is spanned by functions
\be
\varphi_{qj}(z)=q^{-2}z^j-(z-1)^jz^{L-3-j}-q^2(1-z)^{L-3-j},
\ee
with $j=0,\ldots L-3$. The analogue to (\ref{f5}) reads,
\be
\P_q(x,z)=q^{-2}(xz+1)^{L-3}-((x+1)z-x)^{L-3}-q^2(x+1-z)^{L-3}.
\ee
The functional relation $\P_q(x,z)=\a^{L-3}\P_q(\x,z)$ is solved by
\bea
q^{-2}(xz+1)&=&-p_1\a((\x+1)z-\x)\nn\\
(x+1)z-x&=&p_2\a q^2(\x+1-z) \\
q^2(x+1-z)&=&-p_3\a q^{-2}(\x z+1),\nn
\eea
with numbers $p_{1,2,3}$, satisfying $p_i^{L-3}=1$. As 3 divides $L-3$
we can choose $p_1=p_2=p_3=q^{-2}$, which yields
the solution (\ref{rel3}) for $\a$ and $\x$.
We conclude,
\be
\dim \H'= \m(L-2)+\i(L-2),
\ee
and for a state $|\Psi\rangle$ containing $M'$ ordinary and
one exceptional momentum,
\be
\dim\H'= \m(L-2+M')+\i(L-2+M').
\ee
Because we have still $M'+1$ independent BA equations, it follows
\be
\dim\H^{\Psi}_1=\dim\H'-M'-1=\frac{2S^z+1}{3}-1,
\ee
with $S^z=L/2-M'-1\equiv -1\ \mod 3$. This can be written as
\be
\dim\H^{\Psi}_1=\m(2S^z-2)+\i(2S^z-2)+1,
\ee
which is formula (\ref{mult3}).
Now we assume $m=2$, the state $|\Psi\rangle$ contains two exceptional
momenta of equal sign and $M'$ ordinary momenta. By an argument, which
parallels the discussion above, we find for $\dim \H'$:
\be
\dim \H'=\m(L-2+M')+\i(L-2+M').
\ee
But now the number of independent BA equation is not
$M'+2$ but $M'+1$, as the two exceptional momenta are indiscernible.
Therefore,
\be
\dim \H^{\Psi}_1=\dim\H'-M'-1=\frac{2S^z+2}{3}-1.
\ee
Now, $S^z\equiv 1\ \mod 3$, and we find again formula (\ref{mult3}),
which is therefore independent of $m$.

Case III) can be treated without recourse to the wavefunction,
by using the $Z_2$ symmetry of the spectrum. Assume the state
$|\Psi\rangle$ contains one momentum $k_q$ and one momentum $-k_q$.
It is therefore degenerate with a state $|\Psi_0\rangle$ (we use the
notation of section III) having $S^z(\Psi_0)=S^z(\Psi)+2$.
Whereas $S^z(\Psi)\equiv -1\ \mod 3$, $S^z(\Psi_0)\equiv 1\ \mod 3$.
We look for the state $|\Psi'\rangle$ in the multiplet of
$|\Psi_0\rangle$ having $S^z(\Psi')=-S^z(\Psi)$ and determine the number
$k$ of cbs, which have to be added to $|\Psi_0\rangle$ to reach
$|\Psi'\rangle$: $S^z(\Psi_0)-3k=S^z(\Psi')$. We find
\be
k=\frac{S^z(\Psi_0)+S^z(\Psi)}{3}=\frac{2S^z(\Psi)+2}{3}.
\ee
Now, by $Z_2$ symmetry,
\be
\dim\H^{\Psi}_1=\dim\H^{\Psi_0}_{k-1}={\dim\H^{\Psi_0}_1\choose k-1},
\ee
where we have assumed (\ref{mult5}).
Because $|\Psi_0\rangle$ contains no exceptional momenta, we
have
\be
\dim\H^{\Psi_0}_1=\m(2S^z(\Psi)+2)+\i(2S^z(\Psi)+2)=\frac{2S^z(\Psi)+2}{3}=k.
\ee
$|\Psi'\rangle$ is therefore uniquely determined and
\be
\dim\H^{\Psi}_1={k\choose k-1}=k=\dim\H^{\Psi_0}_1.
\ee
Because
\be
\m(2S^z(\Psi)-2)+\i(2S^z(\Psi)-2)=\frac{2S^z(\Psi)+2}{3}-2,
\ee
we find formula (\ref{mult4}). For the case with four additional
exceptional momenta, we have
$S^z(\Psi)\equiv 1\ \mod 3$ and $S^z(\Psi_0)\equiv -1\ \mod 3$,
$S^z(\Psi_0)=S^z(\Psi)+4$ and $k=(2S^z(\Psi)+4)/3$.
We have $\dim\H^{\Psi_0}_1=\dim\H^{\Psi}_1=k$ as above and find
because $\m(2S^z(\Psi)-2)+\i(2S^z(\Psi)-2)$ is now $(2S^z(\Psi)-2)/3$
again formula (\ref{mult4}), independent of $m$.
This concludes the derivation of (\ref{mult})
and (\ref{mult2} -- \ref{mult4}).

\subsection{Numerical Results on short chains.}
All results reported below were obtained by direct diagonalization
of the spin chain hamiltonian (\ref{ham}) for $\D=1/2$ in the sectors with
given $z$-component of the total spin and even chain length $L$.

Table 2) gives the multiplicities deg($S^z$) of the
reference state energy in the sectors with $S^z=S^z_{max}-3l$,
and
lengths $L=12, 14$ and 16:\par
\vspace{4mm}
\no
Table 2):\\
\begin{tabular}{|c|c|c|c|c|c|c|}
\hline
$L$ & energy&deg($L/2$) & deg($L/2-3$) & deg($L/2-6$) & deg($L/2-9$)
& deg($L/2-12$)\\
\hline
12 & 3.0&1&4&6&4&1\\
\hline
14 &3.5&1&4&6&4&1\\
\hline
16 &4.0&1&4&6&4&1\\
\hline
\end{tabular}\par
\vspace{4mm}
\no
These multiplicities coincide with formula (\ref{mult5})
for more than one cbs, i.e. with $\dim\H^0_l$,
in the commensurable ($L=12$) as well as in the
incommensurable cases ($L=14,16$).\\
Tables 3) to 10) gives energies and multiplicities deg($S^z$) for parent
states different from the reference state and on chains
with lengths between 8 and 14.\par
\vspace{4mm}
\no
Table 3). $L=8$, $S^z_{max}=3$:\\
\begin{tabular}{|c|c|c|}
\hline
energy & deg(3) &deg(0)\\
\hline
-1.0 & 1&2\\
-0.4142&2&4\\
1.0&2&4\\
2.4142&2&4\\
3.0&1&2\\
\hline
\end{tabular}\par
\vspace{4mm}
\pagebreak
\no
Table 4). $L=8$, $S^z_{max}=2$:\\
\begin{tabular}{|c|c|c|}
\hline
energy&deg(2)&deg(-1)\\
\hline
-3.6389&1&0\\
-2.5615&2&0\\
-2.4142&2&0\\
-1.1579&1&0\\
-0.9318&2&2\\
-0.7320&2&2\\
0.0&3&4\\
0.4142&2&0\\
0.4823&2&2\\
1.0&1&0\\
1.5176&2&2\\
1.5615&2&0\\
2.0&1&2\\
2.7320&2&2\\
2.9318&2&2\\
3.7969&1&0\\
\hline
\end{tabular}
\par
\vspace{4mm}
\no
Table 5). $L=10$, $S^z_{max}=4$:\\
\begin{tabular}{|c|c|c|c|}
\hline
energy&deg(4)&deg(1)&deg(-2)\\
\hline
-0.5&1&2&1\\
-0.1180&2&4&2\\
0.8819&2&4&2\\
2.1180&2&4&2\\
3.1180&2&4&2\\
3.5&1&2&1\\
\hline
\end{tabular}
\par
\vspace{4mm}
\no
Table 6). $L=10$, $S^z_{max}=3$:\\
\begin{tabular}{|c|c|c|c|}
\hline
energy&deg(3)&deg(0)&deg(-3)\\
\hline
-3.2756&1&2&1\\
-1.6318&1&2&1\\
0.0575&2&4&2\\
0.5&3&6&3\\
0.8847&1&2&1\\
1.5&1&2&1\\
4.3607&1&2&1\\
\hline
\end{tabular}
\par
\vspace{4mm}
\pagebreak
\no
Table 7). $L=12$, $S^z_{max}=5$:\\
\begin{tabular}{|c|c|c|c|}
\hline
energy&deg(5)&deg(2)&deg(-1)\\
\hline
0.0&1&4&3\\
0.2679&2&6&4\\
1.0&2&9&12\\
2.0&2&7&6\\
3.0&2&6&6\\
3.7320&2&6&4\\
4.0&1&2&1\\
\hline
\end{tabular}
\par
\vspace{4mm}
\no
Table 8). $L=12$, $S^z_{max}=4$:\\
\begin{tabular}{|c|c|c|c|}
\hline
energy&deg(4)&deg(1)&deg(-2)\\
\hline
0.1389&2&4&2\\
0.2383&1&2&1\\
1.0&5&12&9\\
2.0&1&6&7\\
3.0&2&6&6\\
4.0463&1&2&1\\
4.8983&1&2&1\\
\hline
\end{tabular}
\par
\vspace{4mm}
\no
Table 9). $L=12$, $S^z_{max}=3$:\\
\begin{tabular}{|c|c|c|c|}
\hline
energy&deg(3)&deg(0)&deg(-3)\\
\hline
-5.3650&1&2&1\\
-4.6365&2&4&2\\
-3.6616&1&2&1\\
-1.8709&1&2&1\\
-1.0&1&8&1\\
-0.8904&1&2&1\\
0.6194&1&2&1\\
\hline
\end{tabular}
\par
\vspace{4mm}
\no
Table 10). $L=14$, $S^z_{max}=6$:\\
\begin{tabular}{|c|c|c|c|c|c|}
\hline
energy&deg.(6)&deg(3)&deg(0)&deg(-3)&deg(-6)\\
\hline
0.5&1&4&6&4&1\\
0.6980&2&8&12&8&2\\
1.2530&2&8&12&8&2\\
2.0549&2&8&12&8&2\\
2.9450&2&8&12&8&2\\
3.7469&2&8&12&8&2\\
4.3019&2&8&12&8&2\\
4.5&1&4&6&4&1\\
\hline
\end{tabular}

\end{document}